\documentclass{article}

\PassOptionsToPackage{numbers, compress}{natbib}
\usepackage[final]{nips_2017}

\usepackage[utf8]{inputenc} 
\usepackage[T1]{fontenc}    
\usepackage{hyperref}       
\usepackage{url}            
\usepackage{booktabs}       
\usepackage{amsfonts}       
\usepackage{nicefrac}       
\usepackage{microtype}      
\usepackage{graphicx}
\graphicspath{{figs/}}

\title{Emotional End-to-End Neural Speech synthesizer}

\newcommand*\samethanks[1][\value{footnote}]{\footnotemark[#1]}

\author{
	\begin{tabular}{ccc}
		Younggun Lee\textsuperscript{1}\thanks{Both authors contributed equally to this work.} &
		Azam Rabiee\textsuperscript{2}\samethanks&
		Soo-Young Lee\textsuperscript{1}\\
	\end{tabular}\\
	\textsuperscript{1}The School of Electrical Engineering\\
	Korea Advanced Institute of Science and Technology, Daejeon, Korea \\
	\textsuperscript{2}Department of Computer Science, Dolatabad Branch, Islamic Azad University, Isfahan, Iran\\
	\texttt{ \textsuperscript{1}\{younggunlee, sy-lee\}@kaist.ac.kr, \textsuperscript{2}a.rabiee@iauda.ac.ir } \\
}

\begin{document}

\maketitle

\begin{abstract}
  In this paper, we introduce an emotional speech synthesizer based on the recent end-to-end neural model, named Tacotron. Despite its benefits, we found that the original Tacotron suffers from the exposure bias problem and irregularity of the attention alignment. Later, we address the problem by utilization of context vector and residual connection at recurrent neural networks (RNNs). Our experiments showed that the model could successfully generate speech for given emotion labels.
\end{abstract}

\section{Introduction}

Recently, researchers presented deep neural network models for text-to-speech (TTS) synthesis whose results are comparable to the previous approaches such as concatenative models. Among the neural TTS models, Tacotron~\citep{wang17} has emerged as an end-to-end TTS model that can be trained from scratch on <text, audio> pairs. The generative model of Tacotron is a sequence-to-sequence (seq-to-seq) model~\citep{suts14} with an attention mechanism~\citep{bahd14}. The model has many advantages compared to other state-of-the-art neural TTS models, like WaveNet~\citep{oord16}, Deep Voice~\citep{arik17}, as well as~\citet{taig17}. No need for separately pre-trained subsections, Tacotron produced natural speech samples which achieved competitive mean opinion score (MOS).

In this paper, we report our experiment on a modified version of the Tacotron architecture to generate emotional speech. We also found that Tacotron had a difficulty to generate a middle part of speech. We resolved it by applying modifications that especially aim facilitating information flow in Tacotron.

We will present an emotional Tacotron to generate emotional speech in Section~\ref{s:emotiontts}. We refer readers to ~\citet{wang17} for details of Tacotron. Section~\ref{s:experiments} describes our experiments. Then, we discuss shortcomings of Tacotron and our solutions in Section~\ref{s:improvements}. The conclusion is given in Sections~\ref{s:conclusion}.

\section{Emotional speech synthesizer}
\label{s:emotiontts}

\begin{figure}[t]
	\centering
	\includegraphics[scale=0.4]{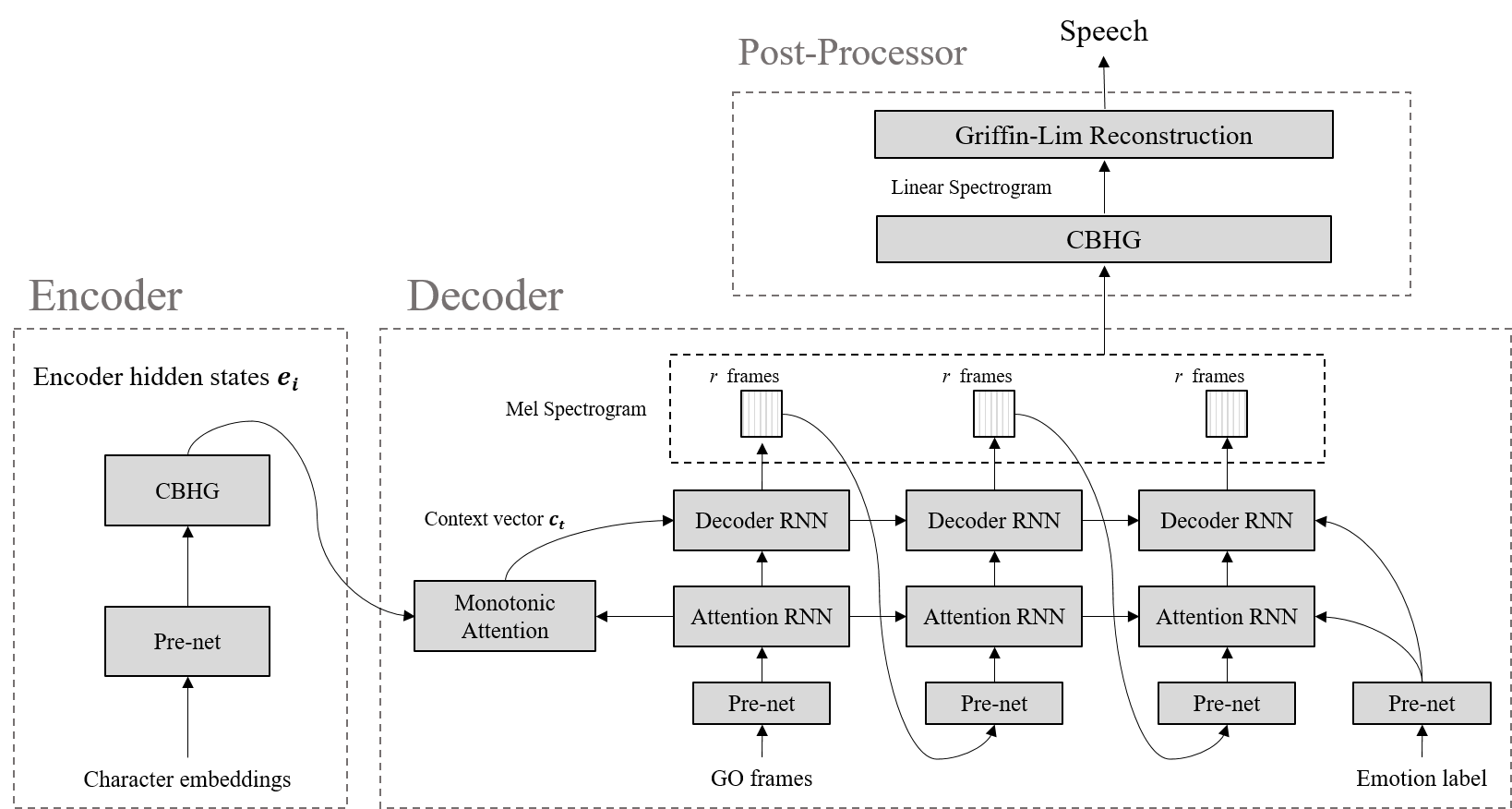}
	\caption{Emotional end-to-end speech synthesizer.}
	\label{fig1}
\end{figure}

As a seq-to-seq model, Tacotron contains three parts: 1) an encoder to extract features from the input text, 2) an attention-based decoder to generate Mel spectrogram frames from the attended portion of the input text, and 3) a post-processor to synthesize the waveform of speech. We present the emotional Tacotron to generate speech that carries special specifications such as emotion or personality, so that the model can have variation in the synthesized speech. We implemented emotional Tacotron by injecting a learned emotion embedding $e$ as follows:
\begin{equation}\label{eq0}
h^{att}_t=AttentionRNN(x_t, h^{att}_{t-1},e) \quad\&\quad h^{dec}_t=DecoderRNN(c_t,h^{dec}_{t-1},e)
\end{equation}
where $x_t$, $c_t$, $h^{att}_t$ and $h^{dec}_t$ are the input, the attention applied context vector, the hidden state of the attention RNN and the hidden state of the decoder RNN at time-step $t$, respectively. Figure~\ref{fig1} depicts architecture of the emotional Tacotron.

\begin{figure}[b]
\begin{center}
	\begin{tabular}{ccc}
		\parbox[c]{4cm}{
			\includegraphics[width=4cm]{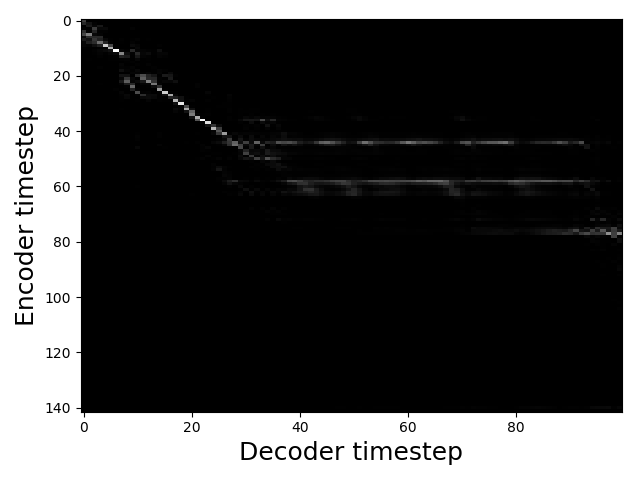}
		}&
		\parbox[c]{4cm}{
			\includegraphics[width=4cm]{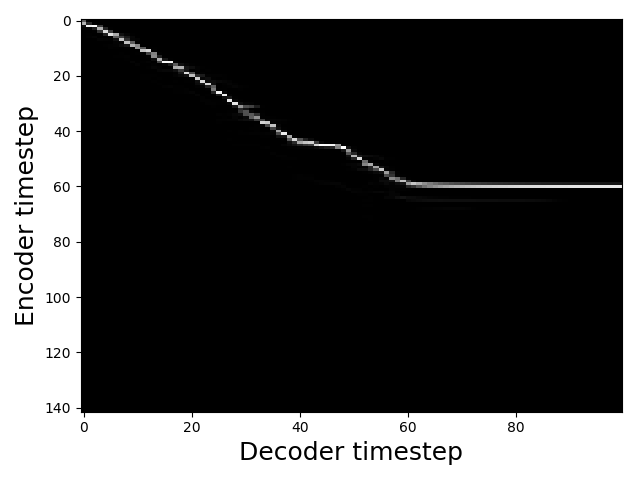}
		}&
		\parbox[c]{4cm}{
			\includegraphics[width=4cm]{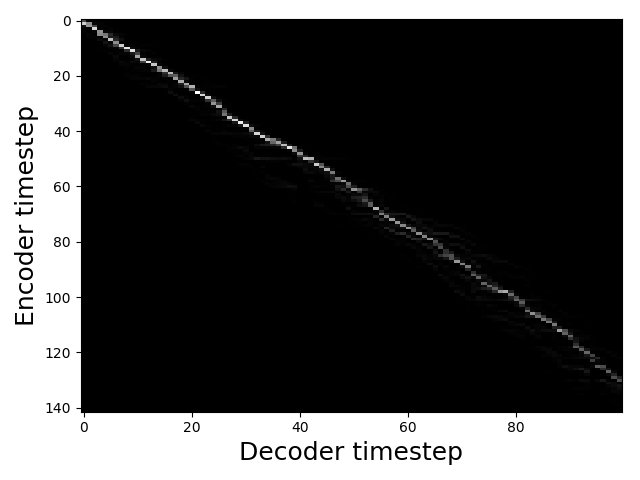}
		}\\
		(a)&(b)&(c)\\
	\end{tabular}
\end{center}
\caption{Attention alignment of a moderate-length sentence (141 characters). (a) Exposure bias problem in the original Tacotron. (b) improvement by monotonic attention (c) improvement by monotonic attention and semi-teacher force training }
\label{fig2} 
\end{figure}

\subsection{Attention-based emotional decoder}

A stack of RNNs is used for the decoder together with a content-based tanh attention mechanism~\cite{taig17} in the original Tacotron. The decoder predicts r-frames of Mel spectrogram at every time-step employing a decoder pre-net, one layer attention RNN, as well as two layer residual connection decoder RNN. For every minibatch, the decoder starts with a “GO” frame which is all zero; later, for every time-step, the previous time frame is fed as the decoder pre-net’s input.

Having inspired by~\citet{tjan17}, we injected the projection of the one-hot emotion label vector to attention RNN by concatenating with the pre-net output and adding one more layer to project it to match the size of the attention RNN input. Same injection is performed at the first layer of the decoder RNN to add the emotional features to the generated spectrogram.

In the original Tacotron, the decoder is trained such that to predict one output at a time by feeding the ground truth information of the previous time-step. However, the ground truth information is not available in the test phase. Hence, the generated output contains some noise for every time-step. As a result, the errors are quickly accumulated and mess the generated wave especially for long wave outputs. The so-called exposure bias problem causes disconnectivity and losing the attention alignment, which shows the messy pattern between decoder time-steps and encoder states. As an example, Figure~\ref{fig2}-(a) shows the attention alignment of the original Tacotron, in which $x$ and $y$ axes belong to the decoder time-step and the encoder time-step, respectively. As can be seen from the figure, the attention works well for the initial time-steps but the model loses the attention at around $t_{decoder}=30$. We addressed the exposure bias problem by the tricks explained as follows.

\paragraph{Monotonic attention (MA). }

Text to speech conversion is basically a monotonic transform. One possible way of handling the non-monotonicity shown in Figure~\ref{fig2}-(a) is forcing the attention to follow a monotonic pattern. Recently,~\citet{raff17} have presented a differentiable method for this purpose. We have implemented this idea in the emotional Tacotron. Figure~\ref{fig2}-(b) shows the attention alignment in the monotonic-attention-added Tacotron, which depicts a clean pattern compared to Figure~\ref{fig2}-(a). Furthermore, having listened to the generated outputs, we have figured out that there is correlation between the quality of the generated output and the cleanness and sharpness of the attention alignment.

\paragraph{Semi-teacher-forced Training (STFT). }

In the conventional teacher-forced training of seq-to-seq model, the decoder's input at each time-step $t$ is the ground truth output (spectrogram frame in our case) at the previous time-step $y_{t-1}$. Feeding ideal input to the decoder in training phase provokes exposure bias problem because noisy generated output $\hat{y}_{t-1}$ from the previous time-step is used in test phase. Having inspired from~\citet{taig17}, we decided to add noise in the training phase by feeding the average of $y_{t-1}$ and $\hat{y}_{t-1}$ to the decoder as $0.5(y_{t-1}+\hat{y}_{t-1})$.

\section{Experiments }
\label{s:experiments}

We trained our emotional Tacotron on the Korean dataset from Acriil that contains <text, audio, emotion label> pairs. A female actor read news articles in six different emotions: neutral, angry, fear, happy, sad, and surprise. The scripts contain same text except for happy, and we observed that different script does not affect the generated happy speech.

We used sequences of less than 200 characters and wav files of shorter than 8.7s after silence trimming. We trimmed silence from the wav files using the WebRTC Voice Activity Detector~\citep{webrtc} because we found that silence trimming is crucial for training. After this process, the dataset contains 21 hours of speech. We did not use masking for the padded zeros at the end of sentences to let the model to train the ending silence. We borrowed the details of the original Tacotron for the common part of our model. In addition, we projected the six categories of the one-hot emotion vector using a fully connected layer with 64 hidden units and 0.5 dropout ratio. 

Since at this moment, we do not have access to a suitable emotion classifier for Korean language and MOS testers, we uploaded our generated speech on Github (\url{https://github.com/AzamRabiee/Emotional-TTS}) instead of quantitative results.

\section{Discussion on prediction of attention alignment }
\label{s:improvements}

When we train the original Tacotron on a non-emotional dataset with same pre-processing, the model had a problem to generate speech. We used a Korean dataset from ETRI, which contains 17 hours of drama scripts recorded by a female actor with natural tone. With the model trained on this dataset, the attention alignment showed irregularities in the middle part. Based on our conjecture that there is correlation between sharpness of attention alignment and the quality of the generated speech, we decided to improve prediction of attention alignment and paid attention to the flow of information in the Tacotron. There are two sources of information for predicting attention alignment. One comes from the hidden state of attention RNN and the other comes from the text encodings from the encoder. Based on these two sources, we propose the following ideas for improving the alignments.

\subsection{Utilization of Context Vector}
\label{ucv}

The attention part in the original Tacotron model often attends to similar part of a text input for several decoder time-steps because pronunciation of one character usually requires more than one frame of spectrogram. Even when the model should change the attention weight, the next part to be attended will be an adjacent part of the currently attended text. Therefore, when deciding attention weights, the model can benefit from having information of previously attended text $c_{t-1}$, which is a weighted sum of text encodings. However, the original Tacotron does not utilize that information; attention RNN takes only spectrogram of the previous time-step, which hardly contain the information of $c_{t-1}$. Based on this idea, we concatenated $c_{t-1}$ to the attention RNN's input $x_t$. Now, the attention RNN takes one more input $c_{t-1}$ as follows:
\begin{equation}\label{eq1}
h^{att}_t=AttentionRNN(x_t,h^{att}_{t-1},c_{t-1}).
\end{equation}

\subsection{Residual Connections in CBHG}

\begin{figure}[t]
	\begin{center}
		\begin{tabular}{cc}
			\parbox[c]{6cm}{
				\centering
				\includegraphics[width=4cm]{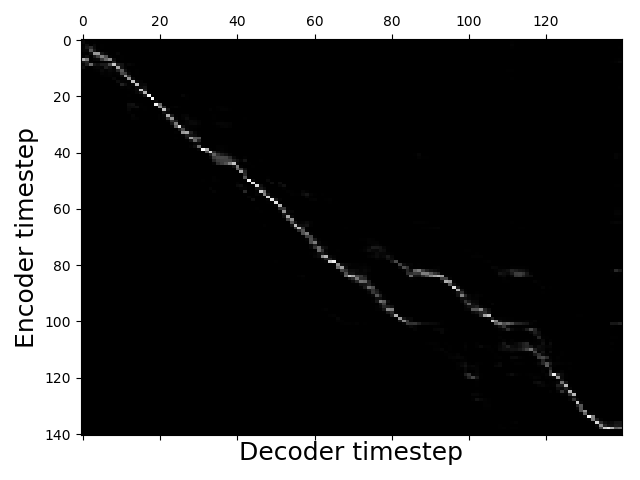}
			}&
			\parbox[c]{6cm}{
				\centering
				\includegraphics[width=4cm]{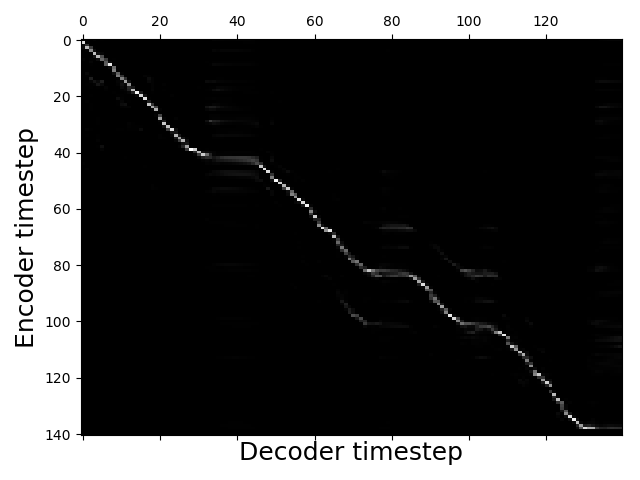}
			}\\
			(a)&(b)\\
		\end{tabular}
	\end{center}
	\caption{Attention alignment of a moderate-length sentence (138 characters). (a) From the original Tacotron. (b) From the Tacotron with utilizing context vector and residual connection
	}
	\label{fig3} 
\end{figure}

For the second idea to improve attention prediction, we change the way of encoding a text input, which is also a source of information for deciding attention alignment. The encoding is generated by the CBHG module (Convolution Bank + Highway + bi-GRU)~\citep{lee16} shown in Figure~\ref{fig1}. The CBHG contains convolutional filter banks, in which each filter explicitly extracts local and contextual information. At the last stage of the CBHG, there is a bi-directional RNN to capture long-term dependency in the text input. Long-term dependency is accumulated in the hidden state of RNN as the RNN reads the input sequentially. One problem here is that the size of hidden state is fixed. If a sequence is long enough, the hidden state will not be able to contain whole information of the sequence. Furthermore, the bi-directional RNN in the CBHG must contain information of current time-step’s text input as well as the information of long-term dependency. This puts more of the burden on the hidden state. In Figure~\ref{fig3}-(a), attention alignments have gaps or blurry parts in the middle of the sequence when the input sentence is long in the original Tacotron. We speculate these irregularities comes from the insufficient capacity of the bi-directional RNN’s hidden state in the CBHG; hence, we decided to add a residual connection that connects the input and the output of the bi-directional RNN. We changed the output of the CBHG to have additional term $x_t$ as follows:
\begin{equation}\label{eq2}
y_t=x_t+BiRNN(x_t,h_{t-1})
\end{equation}
where $x_t$, $h_t$ and $y_t$ are input, hidden state and output of the bi-directional RNN. Now, the residual connection conveys the information of current time-step, so the hidden state of the bi-directional RNN does not need to contain information of current time-step. This connection made the hidden state less congested and helped encoding text information. 

By applying residual connection with the method proposed in Section~\ref{ucv}, we observed sharper and clearer attention alignments as shown in Figure~\ref{fig3}. More examples of alignments and results can also be found on our Github page.

\section{Conclusion }
\label{s:conclusion}

We have proposed a modified Tacotron, as an end-to-end emotional speech synthesizer that takes the character sequence and the desired emotion as input and generates the corresponding wave signal. In our experiments, we have figured out that the quality of the generated speech is highly correlated with the sharpness and the cleanness of the attention alignment; hence, we presented some tricks to improve the attention alignments and therefore quality of the generated wave. We are still investigating more improvement in the model by either improving the quality or speeding up generation. Currently, the model does not care about the phase of the spectrogram which may affect the intelligibility of the emotional wave~\citep{brou17}. On the other hand, the synthesizing process suffers from the low speed of the Griffin-Lim reconstruction. Furthermore, our final goal is to produce a dynamic TTS that is capable of generating speech with different personalities.

\subsubsection*{Acknowledgments}

This work was supported by Institute for Information \& communications Technology Promotion(IITP) grant funded by the Korea government(MSIT) [2016-0-00562(R0124-16-0002), Emotional Intelligence Technology to Infer Human Emotion and Carry on Dialogue Accordingly]

\small

\bibliography{nips_2017}

\begin{thebibliography}{11}
\providecommand{\natexlab}[1]{#1}
\providecommand{\url}[1]{\texttt{#1}}
\expandafter\ifx\csname urlstyle\endcsname\relax
  \providecommand{\doi}[1]{doi: #1}\else
  \providecommand{\doi}{doi: \begingroup \urlstyle{rm}\Url}\fi

\bibitem[Ar{\i}k et~al.(2017)Ar{\i}k, Chrzanowski, Coates, Diamos, Gibiansky,
  Kang, Li, Miller, Ng, Raiman, Sengupta, and Shoeybi]{arik17}
S.~{\"O}. Ar{\i}k, M.~Chrzanowski, A.~Coates, G.~Diamos, A.~Gibiansky, Y.~Kang,
  X.~Li, J.~Miller, A.~Ng, J.~Raiman, S.~Sengupta, and M.~Shoeybi.
\newblock Deep voice: Real-time neural text-to-speech.
\newblock In D.~Precup and Y.~W. Teh, editors, \emph{Proceedings of the 34th
  International Conference on Machine Learning}, volume~70 of \emph{Proceedings
  of Machine Learning Research}, pages 195--204, International Convention
  Centre, Sydney, Australia, 06--11 Aug 2017. PMLR.
\newblock URL \url{http://proceedings.mlr.press/v70/arik17a.html}.

\bibitem[Bahdanau et~al.(2014)Bahdanau, Cho, and Bengio]{bahd14}
D.~Bahdanau, K.~Cho, and Y.~Bengio.
\newblock Neural machine translation by jointly learning to align and
  translate.
\newblock \emph{arXiv e-prints}, abs/1409.0473, Sept. 2014.
\newblock URL \url{https://arxiv.org/abs/1409.0473}.

\bibitem[Broussard et~al.(2017)Broussard, Hickok, and Saberi]{brou17}
S.~Broussard, G.~Hickok, and K.~Saberi.
\newblock Robustness of speech intelligibility at moderate levels of spectral
  degradation.
\newblock \emph{PloS one}, 12\penalty0 (7):\penalty0 e0180734, 2017.

\bibitem[Google()]{webrtc}
Google.
\newblock \emph{WebRTC Voice Activity Detector}.
\newblock URL \url{https://webrtc.org/}.

\bibitem[Lee et~al.(2016)Lee, Cho, and Hofmann]{lee16}
J.~Lee, K.~Cho, and T.~Hofmann.
\newblock Fully character-level neural machine translation without explicit
  segmentation.
\newblock \emph{CoRR}, abs/1610.03017, 2016.
\newblock URL \url{http://arxiv.org/abs/1610.03017}.

\bibitem[Raffel et~al.(2017)Raffel, Luong, Liu, Weiss, and Eck]{raff17}
C.~Raffel, M.~Luong, P.~J. Liu, R.~J. Weiss, and D.~Eck.
\newblock Online and linear-time attention by enforcing monotonic alignments.
\newblock In \emph{Proceedings of the 34th International Conference on Machine
  Learning, {ICML} 2017, Sydney, NSW, Australia, 6-11 August 2017}, pages
  2837--2846, 2017.
\newblock URL \url{http://proceedings.mlr.press/v70/raffel17a.html}.

\bibitem[Sutskever et~al.(2014)Sutskever, Vinyals, and Le]{suts14}
I.~Sutskever, O.~Vinyals, and Q.~V. Le.
\newblock Sequence to sequence learning with neural networks.
\newblock In \emph{Advances in neural information processing systems}, pages
  3104--3112, 2014.

\bibitem[Taigman et~al.(2017)Taigman, Wolf, Polyak, and Nachmani]{taig17}
Y.~Taigman, L.~Wolf, A.~Polyak, and E.~Nachmani.
\newblock Voice synthesis for in-the-wild speakers via a phonological loop.
\newblock \emph{CoRR}, abs/1707.06588, 2017.
\newblock URL \url{http://arxiv.org/abs/1707.06588}.

\bibitem[Tjandra et~al.(2017)Tjandra, Sakti, and Nakamura]{tjan17}
A.~Tjandra, S.~Sakti, and S.~Nakamura.
\newblock Listening while speaking: Speech chain by deep learning.
\newblock \emph{CoRR}, abs/1707.04879, 2017.
\newblock URL \url{http://arxiv.org/abs/1707.04879}.

\bibitem[van~den Oord et~al.(2016)van~den Oord, Dieleman, Zen, Simonyan,
  Vinyals, Graves, Kalchbrenner, Senior, and Kavukcuoglu]{oord16}
A.~van~den Oord, S.~Dieleman, H.~Zen, K.~Simonyan, O.~Vinyals, A.~Graves,
  N.~Kalchbrenner, A.~Senior, and K.~Kavukcuoglu.
\newblock Wavenet: A generative model for raw audio.
\newblock In \emph{Arxiv}, 2016.
\newblock URL \url{https://arxiv.org/abs/1609.03499}.

\bibitem[Wang et~al.(2017)Wang, Skerry-Ryan, Stanton, Wu, Weiss, Jaitly, Yang,
  Xiao, Chen, Bengio, Le, Agiomyrgiannakis, Clark, and Saurous]{wang17}
Y.~Wang, R.~Skerry-Ryan, D.~Stanton, Y.~Wu, R.~J. Weiss, N.~Jaitly, Z.~Yang,
  Y.~Xiao, Z.~Chen, S.~Bengio, Q.~Le, Y.~Agiomyrgiannakis, R.~Clark, and R.~A.
  Saurous.
\newblock Tacotron: Towards end-to-end speech synthesis.
\newblock 2017.
\newblock URL \url{https://arxiv.org/abs/1703.10135}.

\end{thebibliography}
\bibliographystyle{plainnat}

\end{document}